\def\mum{\mu {\rm\,m}}
\def\arcsec{''\hskip-3pt }
\def\arcmin{'\hskip-3pt }

\def\deg{{^{\circ}}}
\def\mjysr{\rm\,MJy{\hskip 3pt}sr^{-1}}

\documentstyle{l-aa}
\input epsf

\begin{document}
 \thesaurus{11(11.09.1 NGC\,6946; 11.09.4, 11.19.2, 11.19.6), 13(13.09.1, 13.09.3)}
\title{ISOCAM\thanks{Based on observations with ISO, an 
	ESA project with instruments funded by ESA Member States 
	(especially the PI countries: France, Germany, the Netherlands
	and the United Kingdom) and with the participation of ISAS 
	and NASA.}  Observations of NGC~6946:  Mid-IR Structure}

\author{S. Malhotra, \inst{1}
\and G. Helou\inst{1} 
\and D. Van Buren\inst{1} 
\and M. Kong\inst{1} 
\and C. A. Beichman \inst{1}
\and H. Dinerstein \inst{2}
\and D. J. Hollenbach \inst{3}
\and D. A. Hunter \inst{4}
\and K. Y. Lo \inst{5}
\and S. D. Lord \inst{1}
\and N. Y. Lu \inst{1}
\and R. H. Rubin \inst{3}
\and G. J. Stacey \inst{6}
\and H. A. Thronson Jr. \inst{7}
\and M. W. Werner \inst{8}
}

   \offprints{san@ipac.caltech.edu}

\institute{IPAC, California Institute of Technology, MS 100-22, Pasadena, CA 91125
\and University of Texas, Astronomy Department, RLM 15.308, Texas, Austin, TX 78712
\and NASA/Ames Research Center, MS 245-6, Moffett Field, CA 94035
\and Lowell Observatory, 1400 Mars Hill Rd., Flagstaff, AZ 86001
\and University of Illinois, Astronomy Department, 1002 W. Green St., Urbana, IL 61801
\and Cornell University,  Astronomy Department, 220 Space Science
Building, Ithaca, NY 14853
\and University of Wyoming, Wyoming Infrared Observatory, Laramie, WY, 82071
\and Jet Propulsion Laboratory, MS 233-303, 4800 Oak Grove Rd., Pasadena, CA 91109}

   \date{Received August 7 1996; accepted 19th August}
   \maketitle
	\markboth {Mid-IR structure of NGC~6946}{Malhotra et al.}

   \begin{abstract}

The nearby spiral galaxy NGC~6946 was observed with ISO-CAM in the
mid-infrared, achieving $7\arcsec$ resolution and sub-$\mjysr$
sensitivity. Images taken with CAM filters LW2 ($7\mu$m) and LW3
($15\mu$m) are analysed to determine the morphology of this galaxy and
understand better the emission mechanisms.  The mid-infrared emission
follows an exponential disk with a scale length $75 \arcsec$. This is
60 \% of the scale length in the optical R-band and radio continuum.
The nuclear starburst region is too bright for reliable measurement in
these images.  Its surface brightness exceeds the inner disk by at
least a factor of 12.  The arms and interarm regions are clearly
detected, with each of these components contributing about equally to
the disk emission.  The arm-interarm contrast is 2-4 in the mid-IR,
close to that measured in the visible R band light and lower than the
contrast in $H\alpha$, suggesting that non-ionizing radiation
contributes significantly to dust heating.

   \end{abstract}

\section{Introduction}

NGC~6946 is a nearby face-on spiral galaxy, a good target for a study
of star-formation and related processes in ``normal'' galaxies. In
addition to the well-resolved disk, it has a starburst nucleus ({\it
e.g.} Engelbracht et al. 1996), and a bright northern spiral arm that
earned it an entry in the Atlas of Peculiar Galaxies (Arp 1966).
NGC~6946 was therefore selected as a focus of the Key Project on the
interstellar medium of normal galaxies (Helou et al. 1996) under NASA
Guaranteed Time on ISO (Kessler et al. 1996).  We report here on the
ISO-CAM (C\'esarsky et al. 1996) maps obtained for this galaxy,
revealing for the first time the mid-infrared (mid-IR) morphology at
vastly improved resolution and sensitivity.

Mid-IR emission in the 5 to 20 $\mum$ range is dominated by very small
grains fluctuating to high temperatures and from Poly Aromatic
Hydrocarbons (PAH) features. Even though these grains are not in
thermal equilibrium, they still convert heating photons (Draine and
Anderson 1985), and should therefore trace star-formation, allowing a
detailed and un-extincted view of that activity.  By studying the
distribution of the mid-IR emission with respect to other components of
the galaxy, i.e. HI, ${\rm H_2}$, ionized gas, and starlight we can also
characterize the heating sources for the mid-IR, and better define its
diagnostic value.

\section{Observations and Data reduction}

NGC~6946 was mapped at $7 \mum$ (LW2 filter, $\Delta \lambda =3.5
\mum$) and at 15 $\mum$ (LW3 filter, $\Delta \lambda = 6 \mum$), using
the raster scan mode to cover roughly $ 12.5 \arcmin \times 12.5
\arcmin$ centered on the nucleus.  CAM was set to $6 \arcsec$/pixel,
and the raster was made up of $8 \times 8$ pointings separated by $81
\arcsec$, or 13.5 pixels in each direction, allowing 
for better spatial sampling.  At each pointing in the raster scan we
took 8 frames integrating 5 seconds for each at 7 $\mum$, and 10
frames of 2 seconds each at 15 $\mum$. H$\alpha$ and broad-band
imaging of NGC~6946 was done at Palomar observatory 60-inch telescope.

The first step in reducing the ISO-CAM data was deglitching, done
within the ISOCAM Interactive Analysis package using the routine
``mr1d\_deglitch'', found most reliable among the available procedures
for removing cosmic ray hits.  Detector transients were then removed
within the same software package, using the ``IPAC'' transient removal
algorithm, which uses a simplified physical model for each pixel's
time dependent behavior.  Then a set of eight ``local'' flat fields
were constructed by taking medians of sequential galaxy-free frames
and were then used to form interpolated flats for each of the
individual frames, by simple linear interpolation for each pixel.  The
dark-removed, deglitched and transient corrected data were then
flattened by dividing each frame by the appropriate interpolated flat
field. Using these processed data we then constructed a model for the
sky flux distribution in an array with $3 \arcsec \times 3
\arcsec$ pixels.  This finer array is required because the raster
steps amount to 13.5 of the $6 \arcsec \times 6 \arcsec$ camera
pixels.  We assume the raster steps correspond exactly to commanded
values.  A least-squares fit was then made to determine the maximum
likelihood value of the sky's average surface brightness in each
pixel.  This scheme of modelling the sky avoids some of the error
propagation problems of the more traditional means of producing
mosaiced images (cf Van Buren and Kong, 1996).

The diffraction-limited beam of ISO has a FWHM of $3 \arcsec$ at $7
\mum$ and $6.3 \arcsec$ at $15 \mum$, but this resolution is not
achievable with our observing mode. The 7 $\mum$ image contains
foreground stars whose FWHM measures about $7.3 \arcsec$.  At $ 15
\mum$ the stars are very faint, and only one of them has a measurable
FWHM of $7.2 \arcsec$. The noise levels in these images are
approximately $0.1 \mjysr\simeq\rm 2\mu Jy\hskip3pt arcsec^{-2}$,
roughly 3 times lower than the noise in the ``Auto-Analysis Results''
map (OLP4.0).  Features roughly a thousand times brighter are still
reliably measured in these maps.  The inner $15\arcsec$ or so are too
bright to measure, and the surface brightness at the nucleus exceeds
$300$ and 560 $\mjysr$ respectively in LW2 and LW3.

These maps result from a preliminary data reduction which could be
significantly improved, especially in the areas of transient removal
and flat fields.

\begin{figure}[htbp]
\epsfxsize=8.0cm\epsfbox{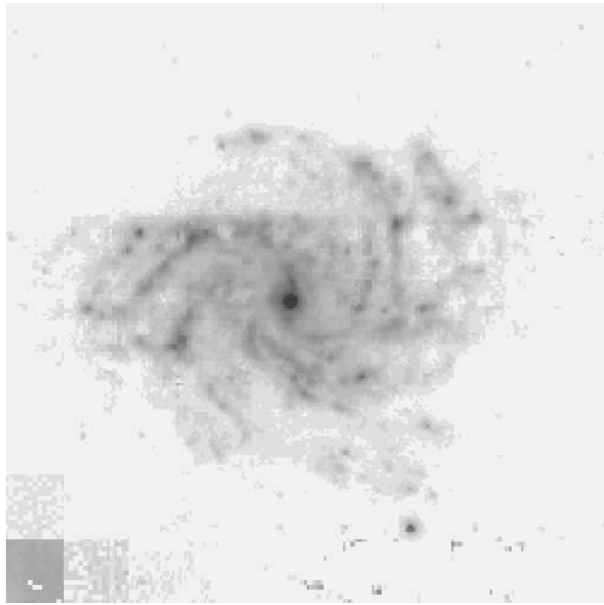}
\caption{ISOCAM image of NGC~6946 in the LW2 filter at a wavelength of
 $7 \mum$, $\Delta \lambda =3.5 \mum$. This image is a mosaic of $ 8
\times 8$ pointings. Log scale of brightness is used to display the dynamic
 range of the exponential disk surface brightness.  Some Galactic
foreground stars are visible to the south and west of NGC
6946. Fortuitous alignment of several features produces what seems to
be an ``edge'' parallel to the major axis in the north. This is at
least partly real and visible in optical images, and is exaggerated by
residual responsivity drifts unaccounted for by the time-dependent
flat field. The image is rotated $21 \deg$ clockwise w.r. to images
showing North at top and east to the left.}
\end{figure}

\begin{figure}[htbp]
\epsfxsize=8.0cm\epsfbox{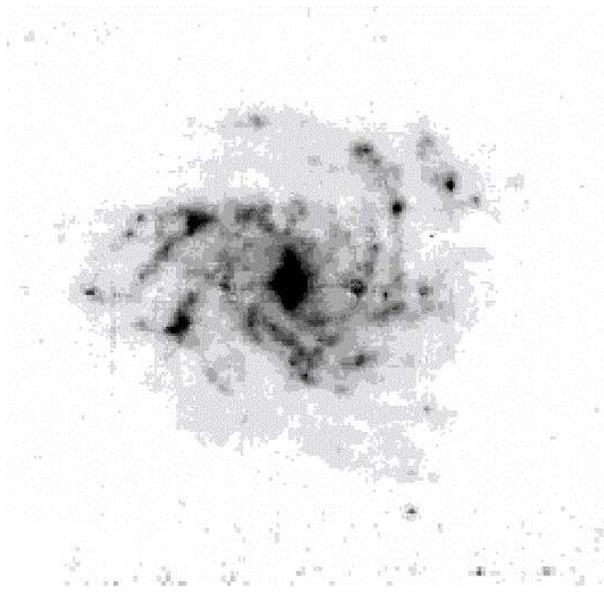}
\caption{ISOCAM image of NGC~6946 in the LW3 filter at a wavelength of
 $15 \mum$, $\Delta \lambda = 6 \mum$ (cf caption of Figure 1 for
details). The image is displayed with a different greyscale stretch
from figure 1 to show the faint outer disk}
\end{figure}

\section{Disk Structure}

The ISOCAM maps of NGC~6946 have enough resolution, sensitivity and
extent to allow a detailed comparison with the maps at other
wavelengths. The mid-IR images at both $ 7 \mum$ and $15 \mum$ show a
good qualitative similarity to optical, radio and $H\alpha$
images. Diffuse emission from the disk in the interarm regions is
clearly detected (Figures 1 and 2) and can be traced out to $\sim 5
\arcmin$, comparable to the optical size $ D_{25}=10-12 \arcmin$.

Except for the region in the inner $50 \arcsec$ of the galaxy where
oval distortions are present (Zaritsky \& Lo 1984), the mid-IR light
follows an exponential profile (Figure 3). The radial profile of the
galaxy is calculated by deprojecting the galaxy to the plane of the
sky (assuming an inclination of $30 \deg$ and position angle of $69
\deg$, subtracting the mean sky from an annulus outside the galaxy and
then taking median emission levels at different annuli. Taking the
median emission in annuli instead of the mean minimizes contributions
from very bright regions of the spiral arms. Both LW2 and LW3 images
were analysed in this manner and show identical scale lengths of $75
\arcsec$ at radii between $70 \arcsec $ and $200 \arcsec$. This scale
length is mildly dependent on range of radii where the fit is taken,
and on the sky subtraction; we estimate the uncertainty at about 10\%.

\begin{figure}[htbp]
\epsfxsize=8.0cm\epsfbox{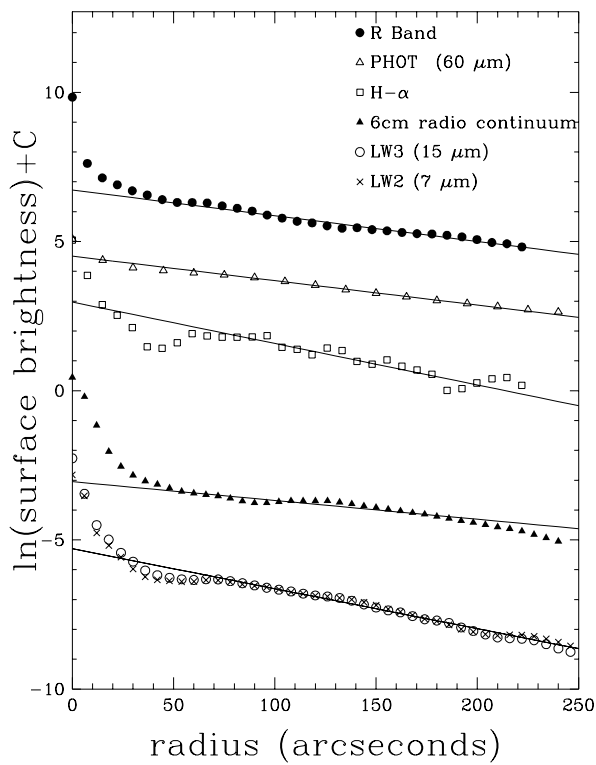}
\caption{The radial profile of NGC~6946 is exponential at most
wavelengths including the mid-Infrared, but with different scale
lengths. We show the profiles in the mid-IR from ISOCAM (7 \& 15 $\mum$)
, radio continuum (6 cm), $H\alpha$, Far-Infrared ($60 \mum$)
and Optical R band.}
\end{figure}

For comparison we derive scale lengths for the optical R-band, and
$H\alpha$ using the same method as for the mid-IR emission.  The
results are listed in Table~1.  Also listed are scale-lengths derived
by previous studies of this galaxy.  We see that the mid-IR emission
and $H\alpha$ emission have similar scale-lengths closest to the scale
length of CO distribution of $86 \arcsec$ (Tacconi \& Young 1986,
TY86). Red light in optical R band shows a scale length of $115
\arcsec$. To reconcile the optical and mid-IR scale lengths we
need to have made in the CAM images a sky subtraction error about ten
times greater than the pixel-to-pixel rms noise.  The ratio of mid-IR
to radio behaves much like the ratio of FIR to radio, falling by the
expected factor of 2 from a radius of $100 \arcsec$ to $200
\arcsec$ (Lu et al. 1996).


\begin{table}
\caption[ ]{Scale lengths}
\begin{flushleft}
\begin{tabular}{lccl}
\hline\noalign{\smallskip}
Wavelength & scale length & range of fit & Source\\
\noalign{\smallskip}
\hline\noalign{\smallskip}
7 $\mum$   	& $74.6 \arcsec$	& $70-200 \arcsec$ & this paper\\
15  $\mum$ 	& $74.6 \arcsec$	& $70-200 \arcsec$ & this paper\\
$H \alpha$ 	& $71.9 \arcsec$	& $70-200 \arcsec$ & this paper\\
R-band (0.70 $\mum$) & $115.9 \arcsec$  & $70-200 \arcsec$ & this paper\\
6 cm continuum  & $ 109-140 \arcsec$	& $90-240 \arcsec$ & Lu96\\
60 $\mum$ 	& $78-115 \arcsec$	& $90-240 \arcsec$ & Lu96 \\
HI (21 cm)      & $294 \arcsec$		& $200-580 \arcsec$ & TY86\\
$H_2$		& $86 \arcsec$		& $40-280  \arcsec$ & TY86 \\
Total ISM (HI+$H_2$) & $ 112 \arcsec$	& $40-280  \arcsec$ & TY86 \\
\noalign{\smallskip}
\hline
\end{tabular}
\end{flushleft}
\end{table}

\section  {Arm-Interarm Contrast}

A striking feature of the ISO-CAM maps of NGC~6946 is the conspicuous
diffuse disk underlying the arms.  The disk is relatively symmetric,
making it unlikely to be the result of transient effects in detectors
or other artifacts.  In a semi-flocculent galaxy like NGC~6946 it is
difficult to quantify the arm-interarm contrast.  We define this
contrast as the ratio between the peak in the arms and the median
value of brightness in an annulus at the same radius.  The arm regions
were selected from the maps by eye. The contrast varies between 2 and
4 with the northern arm showing a higher arm brightness. The
arm-interarm contrast rises from 2 to 4 with increasing radius along
the northern arm (Figure 4). The southern arm is not continuous and
shows arm-interarm contrast between 1 and 2 in the inner parts of the
galaxy and up to $\sim$4 in the outer section which is disjointed
(Figures 1 and 2). These results are consistent with TY86 who find
that the arm-interarm ratio in CO emission increases with radius.

Comparison of the arm-interarm contrast between mid-IR and $H\alpha$
emission could address the question of what heats the dust, and more
specifically whether it is the ionizing radiation which is responsible for
the mid-IR emission (Devereux \& Young 1993) or whether diffuse soft
UV and visible starlight plays a significant role (Persson \& Helou
1987). If the dust heating is primarily from ionizing radiation, the
mid-IR emission should follow the distribution of ionized gas as
traced by H$\alpha$ for example. The data in hand (Figure 4) suggest that
the arm-interarm contrast is higher for H$\alpha$ than for the mid-IR.
Since our H$\alpha$ images do not detect interarm emission, our estimate
of the discrepancy in contrasts is only a lower limit.  The H$\alpha$ and
Mid-IR emission have similar scale lengths in the disk as shown in the
previous section, whereas they have dissimilar arm-interarm contrast.  The
arm-interarm contrast in Mid-IR is closer to the contrast in R band light,
which has a much longer scale length than mid-IR or H$\alpha$.  One
explanation may be that the PAHs are destroyed near HII regions in the
arms, thus softening the contrast between arms and interarms.

\begin{figure}[htbp]
\epsfxsize=8.0cm\epsfbox{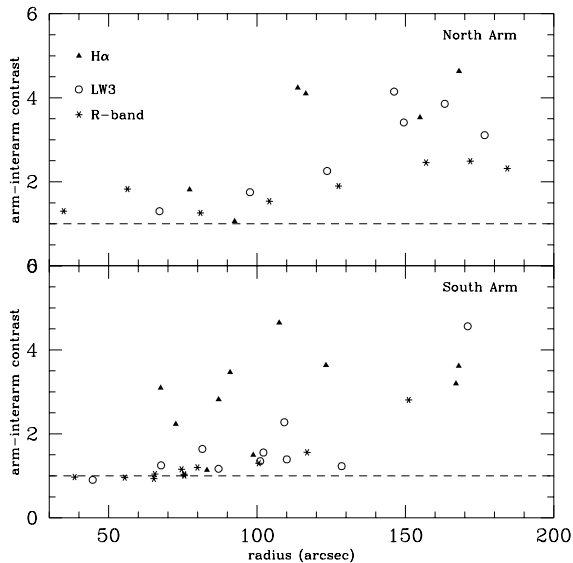}
\caption{The Arm-Interarm contrast (defined in \S4) is plotted as a 
function of radius for the northern and the southern arms. The
triangles represent the arm-interarm contrast in the $H\alpha$ image
smoothed to the CAM resolution and the open circles show the contrast
for LW3 ($15\mum$) image. The asterisks show the arm-interarm
constrast for smoothed R band images}
\end{figure}

\section{Nucleus}

Various indicators point to NGC~6946 as having a starburst nucleus
(e.g. van der Kruit, Allen \& Rots 1977, Rieke 1976; Telesco \& Harper
1980; 
DeGioia-Eastwood et al. 1984).  The starburst nature of the nucleus is
confirmed by the ISOCAM data, which show a greatly enhanced surface
brightness, 12 and 18 times the surface brightness of the nearby inner
disk at $7 \mum$ and $15 \mum$ respectively (Figure 3). These values
are lower limits because the nucleus is beyond the linear regime of
the CAM detector in these data. Telesco, Dressel \& Wolstencroft (1993)
measure the peak brightness at 10.8 $\mum$ to be 0.4 Jy. The size of
the nucleus is measured to be 100 pc at 10 $\mum$ (Telesco et
al. 1993) and $8 \arcsec$ ($\sim$ 200 pc) at 6 cm continuum Beck \&
Hoernes (1996).

\section{Conclusion}

Based on this preliminary reduction of ISO-CAM data, the mid-IR
morphology of NGC~6946 is quite distinct from other traditional
tracers of the interstellar medium, and a promising addition because
it is largely free of extinction.  The maps clearly show the familiar
structure of arms, exponential disk, and high brightness star-burst
nucleus.  However, they also suggest that the mid-IR may be a more
complex tracer - it is closest to H$\alpha$ in disk scale length, but
resembles most the visible R band in arm-interarm contrast.  The
mid-IR emission is not distributed like atomic gas which has a hole in
the center (TY86) nor like total gas distribution, but its radial
profile is similar to that of molecular gas (which resembles
H$\alpha$).  Though puzzling and subject to more thorough data
reduction and analysis, these properties are not necessarily
contradictory, since each of the tracers is a product of different
physical components modified by several effects.

\begin{acknowledgements}
This work was supported  by ISO data analysis funding from  the US
National Aeronautics and Space Administration, and carried out at the
Infrared Processing and Analysis Center and the Jet Propulsion Laboratory
of the California Institute of Technology.

\end{acknowledgements}


\begin{thebibliography}{}
\bibitem{} Arp, H, 1966, ApJS, 14, 1.
\bibitem{} Beck R. \& Hoernes, P., 1996,Nature 251,15
\bibitem{} C\'esarsky et al., 1996, A\&A, this issue.
\bibitem{} DeGioia-Eastwood, K., Grasdalen, G.L., Strom, S.E., Strom, K.M, 1984, ApJ, 278, 564
\bibitem{} Devereux, N.A., Young, J., 1993, AJ 106, 948.
\bibitem{} Draine , B.T., Anderson, N. 1985, ApJ, 292, 494.
\bibitem{} Engelbracht, C.W., Rieke, M, Rieke, G.H., Latter, W., 1996, ApJ in press.
\bibitem{} Helou et al. 1996, A\&A, this issue.
\bibitem{} Kessler et al. 1996, A\&A, this issue.
\bibitem{} Lonsdale Persson, C.J., Helou, G.X., 1987, 314, 514
\bibitem{} Lu N et al., 1996, A\&A, this issue.
\bibitem{} Rieke, G.H., 1976, ApJ, 206, L15	
\bibitem{} Tacconi, L., Young, J., 1986, ApJ, 308, 600.
\bibitem{} Telesco, C.M., Dressel, L.L., Wolstencroft, R.D., 1993, ApJ 86, 286
\bibitem{} Telesco, C.M., Harper, D.A., 1980, ApJ, 235,392.
\bibitem{} Van Buren, D., Kong, M., 1996, in preparation. 
\bibitem{} van der Kruit, P., Allen, R., Rots A.H. 1977, A\&A, 55, 421.
\bibitem{} Zaritsky, D., Lo, K.Y., 1986, ApJ, 303, 66.
\end{thebibliography}
\end{document}